# What is the effect of synergy in international collaboration on regional economies?

Inga Ivanova,[1] Øivind Strand,[2] & Loet Leydesdorff[3]

**Abstract**

We analyze the effects of relative increments of mutual information among the geographical, technological, and organizational distributions of firms on the relative augmentation of regional summary turnover in terms of synergies. How do increases in synergy in international cooperation affect regional turnover? The methodological contribution of this study is that we translate the synergy (abstractly measured in bits of information) into more familiar economic terms, such as turnover for the special case of domestic-foreign collaborations. The analysis is based on Norwegian data, as Norway is a small country with an open and export-oriented economy. Data for Norway is publicly available in great detail.

## Introduction

Recent advances in Triple Helix (TH) theory regarding indicators for synergy in innovation systems [see Leydesdorff and Park, 2014 for a review] open possibilities to research how dimensions like local/global in firm-level data influence TH synergy. Lengyel and Leydesdorff (2011) showed a weak correlation between international ownership and synergy in the regional innovation systems of Hungary. This study compares data of the 500 largest firms (in terms of turnover) in two Norwegian counties: one industrial county with high levels of synergy and weak academic institutions, and another with low synergy but strong academic institutions.

---

[1] corresponding author; Institute for Statistical Studies and Economics of Knowledge, National Research University Higher School of Economics (NRU HSE), 20 Myasnitskaya St., Moscow, 101000, Russia inga.ivanova@hse.ru;

[2] Norvegian University of Science and Technology (NTNU) Aalesund, Department of International Business, PO Box 1517, 6025 Aalesund, Norway; +47 70 16 12 00; ost@hials.no

[3] Amsterdam School of Communication Research (ASCoR), University of Amsterdam, PO Box 15793, 1001 NG Amsterdam, The Netherlands; loet@leydesdorff.net

The crisis in the offshore-related industries caused by the dramatic drop in the oil prices in the autumn of 2014 triggered a re-structuring of the industry: the effects of foreign ownership and re-design of the global value chains are high on the political agenda in Norway. A possible way to enhance the quality and efficiency of a regional economy lies in the development of both regional and cross-border collaborations. The central role of such collaborative interaction has been discussed in both the cluster [Porter 1998; 2000, Ketels, 2011] and the global value chain (GVC) literature [Humphrey and Schmitz, 2000; 2002; Gereffi et al. 2005]. The former emphasizes interactions between local firms and knowledge institutions, whereas the latter gives prime importance to interaction with global buyers. Collaborations can provide added value as a result of the creation of new products and market services. GVC literature is mainly concerned with governance and upgrading of the global value chains [Gereffi and Lee, 2012]. Governance is the coordination of economic activities through non-market activities [Humphrey and Schmitz, 2002] and upgrading refers to shift of activities due to increasing competitive pressure. These authors also stress that "governance is particularly important for the generation, transfer and diffusion of knowledge leading to innovation." In the governance of GVCs, the lead firms play an important role as described in detail by Gereffi and Lee [2012].

The core of a region's economic success is dependent on the quality of its innovation system and the ability for firms located in the region to sustain competitive advantages [Maskell and Malmberg, 1999 a, b] in an economy dominated by global value chains [Gereffi and Lee, 2012; 2016]. The concept of regional innovation systems [Cooke, 1992] was articulated in reaction to the concept of national systems of innovations [Freeman, 1987; Lundvall, 1988; 1992], but is relatively new as a metaphor used at the level of policy making [Cooke & Memedovic, 2003]. The relations between clusters, regional innovation systems (RIS) and the global economy have been investigated by Asheim et al., 2006. These authors emphasize that the main elements of a RIS are the regional production structures and supportive infrastructure with knowledge institutions as the most important element. European cluster survey [Isaksen 2005] found an increasing number of firms sourcing outside the cluster and an increased presence of multinational companies (MNCs) in the clusters. In a study of knowledge-brokers in RIS,

Kauffeld-Monz and Fritsch (2013) found that this role is taken by local knowledge institutions but also by MNC's, especially in lagging regions.

Systems of innovations can be analyzed in terms of the Triple Helix (TH) model of innovations [Leydesdorff & Etzkowitz, 1996]. The TH metaphor links economics, sociology, and innovation theory by studying the network of institutional relations among universities, industries, and governmental agencies. A TH innovation system comprises interactions among three major institutions – science, government, and industry – which are responsible for economic development and knowledge production. In innovative regions, a strong and constant interaction among these actors is assumed. The interactions are especially important for cross-border regions which wish to enhance their innovation performance [Lundquist & Trippl, 2013].

This study addresses the question of how regional development is wired to international collaborations. We offer a theoretical and empirical framework to address this question. The focus here is on the extension of synergy measurement techniques in regional innovation systems, being one of the major developments in the TH literature, with looking at the synergies created within the set of foreign-owned firms. We consider the TH system of relations as an eco-system that can be more or less synergetic relative to the interactions among agents. A measure for synergy is provided by the mutual information in three (or more) dimensions [Yeung, 2008]. Mutual information can be calculated using the TH indicator, which was first developed for interactions among geographical, technological, and organizational distributions [Leydesdorff *et al*., 2006]. The TH indicator is based on information theory and enables us to measure synergy in a TH system of relations in terms of bits of information. However, one may have difficulties understanding what bits of information would mean in familiar economic terms (cf. Theil, 1972), such as the ones used to determine the level of territorial economic development, that is, the aggregate turnover of all the enterprises in a territory. This study is an attempt to answer this question.

The economic potential of regions can be augmented by technological development. Technology transfer can be considered a means for maximizing the potential of technologies. Transferred technologies play the role of complements to the economic structure, socio-economic institutions, and innovative capabilities of regions. Newly acquired technologies create room for new combinations and new markets, and are often considered the main drivers of cross-border collaborations [Van Den Broek & Smulders, 2013].

New technologies can be transferred as a part of foreign participation in domestic firms. Foreign participation is often a result of the emergence of global value chains, established by multinational corporations in order to enhance their profit margin [Gereffi, 1994]. Firms with foreign participation can contribute to regional development. The net value of products and services produced with the transferred technologies can be considered as additional input to cross-border markets. One of the core questions which policy makers responsible for the economic development of regions have to answer can be formulated as follows: should more attention be paid to generating synergy in international collaboration or in the domain of domestic firms? What would enhance efficiency in the development of regional economies? The research question of the present study is an attempt to answer this question on the basis of the TH approach. We analyze two Norwegian regions: Møre og Romsdal and Sør-Trøndelag, since these regions possess a very different economic structure and can be used as examples for demonstrating the possible outcomes, that is, whether international collaborations or domestic firms are of primary importance for regional development.

**Method and data**

Firm-level data on domestic and foreign ownership and their turnover for two Norwegian regions – Møre og Romsdal and Sør-Trøndelag – were constructed based on data from the PureHelp[4] database (on municipality number, NACE code, number of employees, and turnover) and Proff[5] database (for ownership data). The databases were manually matched for the 500 firms with highest turnover in the counties; data for each company was then transferred to Excel files. The choice of the 500 largest firms in the sample was made because these firms provide the major part of aggregate turnover in each of two counties under study. The records include municipality code, NACE code, size code, turnover (in Norwegian Kroner (NOK)), type of ownership, and also international and national turnover for the mother company. A level of at least 20% foreign ownership is used as cutoff for the attribution of foreign ownership.

We use high-level aggregation of the ISIC/NACE categories, listed in Appendix A, [Eurostat, 2008] to differentiate the firms with respect to the technological dimension. The organizational dimension is subdivided into eight classes according to the number of employees: zero employees; 1-4 employees; 5–9 employees; 10–19 employees; 20–49 employees; 50–99; 100–249; >250 employees. Firms with different sizes can be expected to have different organizational structures, business models, and economic dynamics [Blau and Schoenherr, 1971].

---

[4] www.purehelp.no

[5] www.proff.no

Mutual information in three dimensions – geographical, organizational, and technological – at national and regional levels has been calculated for a number of countries, such as the Netherlands [Leydesdorff, Dolfsma, & Van der Panne, 2006], Germany [Leydesdorff & Fritsch, 2006], Hungary [Lengyel & Leydesdorff, 2011], Sweden [Leydesdorff, & Strand, 2013], Norway [Strand & Leydesdorff, 2013], Russia [Leydesdorff, Perevodchikov, & Uvarov, 2015]. The mutual information among three dimensions $T_{GOT}$ can be defined as the ternary interception area of three corresponding dimensions: $H_G$, $H_O$, $H_T$ (Fig.1) which in formula format can be also written as [Abramson, 1963; Ashby, 1964]:

$$T_{GOT} = H_G + H_O + H_T - H_{GO} - H_{GT} - H_{OT} + H_{GOT} \qquad (1)$$

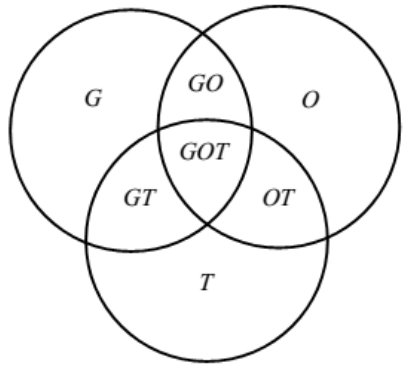

Figure1. Vienn diagram of three intercepting - geographical, organizational, and technological - dimensions.

Indices *G, O, T* in Eq. 1 refer to the geographical, organizational, and technological dimensions. The firms in each of the two counties were grouped with respect to municipalities, according to municipality codes (geographical dimension *G*), number of employees (organizational dimension *O*), and technology class according to NACE codes (technological dimension *T*). The corresponding Shannon entropy measures $H_i, H_{i,j}, H_{i,j,k}$ (here indices *i, j, k* stand for *G, O, T*) can be formulated as follows:

$$H_i = -\sum_i p_i \log p_i$$

$$H_{ij} = -\sum_i p_{ij} \log p_{ij} \qquad (2)$$

$$H_{ijk} = -\sum_i p_{ijk} \log p_{ijk}$$

The probabilities: $p_i, p_{ij}, p_{ijk}$ are defined as the ratio of the number of firms in the corresponding subdivision to the total number of firms in a region. For example, $p_G = \frac{n_G}{N}$, where $n_G$ is the number of firms in the municipality with index *G*, and *N* is the total number of firms in the county, to which the municipality with index *G* belongs, etc. $T_{GOT}$ is a signed information measure [Yeung, 2008] and consequently cannot be considered Shannon-type information [Krippendorff, 2009]. The case when this information measure is negative can be interpreted as reduction of uncertainty that prevails at a regional systems level.

Leydesdorff and Ivanova, [2014] conceptualized $T_{GOT}$ as mutual redundancy originating in positionally differentiated inter-human communication systems. Positional differentiation of communication systems in relation to one another means that systems entertain different sets of communication codes which supply specific meaning to the information. Mutual redundancy measures the surplus of options (that is, redundancy) generated when meaning processing systems communicate in terms of informational exchanges. This surplus of options itself increases the overall uncertainty. However, if the resulting redundancy is negative uncertainty is decreased. The larger this decrease of uncertainty, the more "synergetic" is the interaction among the communicating systems. In other words, negative valued $T_{GOT}$ can also be called the synergy in interactions.

Ternary synergy among geographical, technological, and organizational distributions $T_{GOT}$ was calculated for all the regional firms, including nationally owned and those with foreign participation, and separately for the national firms only ($T_{GOT}^{nat}$). Here synergy for the firms with foreign participation only is defined as a difference between summary synergy and a synergy generated by domestically owned firms:

$$T_{GOT}^{int} = T_{GOT} - T_{GOT}^{nat} \tag{3}$$

where the term $T_{GOT}^{int}$ accounts for synergy formed by purely international firms and national-foreign interactions. We define the contribution of the firms with foreign participation as net input of international dimension plus an interaction term, since the presence of firms with foreign participation generates the interactions:

$$T_{GOT}^{int} = T^{*}{}_{GOT}^{int} + \tilde{T}_{GOT} \tag{4}$$

Here the term $T^{*}{}_{GOT}^{int}$ refers to synergy generated exclusively by firms with foreign participation, while the term $\tilde{T}_{GOT}$ refers to national-foreign interactions. Inputs from national, international, and interaction synergy can thus explicitly be distinguished. For example, $H_G$, which is defined as:

$$H_G = -\sum_G \frac{n_G^{nat}+n_G^{int}}{N} \log \frac{n_G^{nat}+n_G^{int}}{N} \tag{5}$$

can be re-written in the form:

$$H_G = -\sum_G \frac{n_G^{nat}}{N} \log \frac{n_G^{nat}}{N} - \sum_G \frac{n_G^{nat}}{N} \log(1 + \frac{n_G^{int}}{n_G^{nat}}) - \sum_G \frac{n_G^{int}}{N} \log \frac{n_G^{int}}{N}$$
$$- \sum_G \frac{n_G^{int}}{N} \log(1 + \frac{n_G^{nat}}{n_G^{int}})$$

$$= H_G^{nat} + H_G^{int} + \widetilde{H}_G \qquad (6)$$

here $H_G^{nat}$:

$$H_G^{nat} = -\sum_G \frac{n_G^{nat}}{N} \log \frac{n_G^{nat}}{N} \qquad (7)$$

is the input of domestically owned firms, whereas $H_G^{int}$ refers to contribution of firms with foreign participation:

$$H_G^{int} = -\sum_G \frac{n_G^{int}}{N} \log \frac{n_G^{int}}{N} \qquad (8)$$

and $\widetilde{H}_G$:

$$\widetilde{H}_G = -\sum_G \frac{n_G^{nat}}{N} \log(1 + \frac{n_G^{int}}{n_G^{nat}}) - \sum_G \frac{n_G^{int}}{N} \log(1 + \frac{n_G^{nat}}{n_G^{int}}) \qquad (9)$$

accounts for the interaction between national and international dimensions. Analogously, one can distinguish among national, foreign, and interaction inputs for all entropy terms provided in Eq. 2. Correspondingly, the summary synergy can be written as the net inputs of national ($T_{GOT}^{nat}$) and international ($T^{*}{}_{GOT}^{int}$) dimensions plus interaction term ($\widetilde{T}_{GOT}$), as follows:

$$T_{GOT} = T^{*}{}_{GOT}^{int} + \widetilde{T}_{GOT} + T_{GOT}^{nat} \qquad (10)$$

Firms with foreign participation can be attributed to international collaborations. The economic effect of the collaboration can be measured in terms of turnover. The turnover of firms with foreign participation (international turnover $R_{int}$) is a fraction of summary turnover ($R$) produced by both domestic firms and firms with foreign participation. The ratio of international to summary turnover can be interpreted as the share of international collaboration activities in the total regional turnover. One may wonder if there is a link between real economic output, measured in terms of turnover $R$, and the more abstract entropy-based measures, such as the synergy in interactions? In other words, how may the synergy affect the turnover, and what practical implications may this have.

The ratio $T_{GOT}^{int}/T_{GOT}$ as a function of $R_{int}/R$ can vary between 0 and 1 (this is independent of the number of firms in the sample). If there are no firms with foreign participation then all the turnover is generated by domestically owned firms only and $R_{int}/R = 0$. In this case all the synergy is also generated by domestic firms and $T_{GOT}^{int}/T_{GOT} = 0$. If foreign participation is the case for all the firms then both $R_{int}/R$ and $T_{GOT}^{int}/T_{GOT}$ are equal to unity. We expect monotone behavior of $R$ as a function of $T$, that is, sequential increase in the percentage of foreign-owned firms entails a sequential increase in $R_{int}/R$ and increase in $T_{GOT}^{int}/T_{GOT}$, so that $R_{int}/R$ and $T_{GOT}^{int}/T_{GOT}$ vary in the interval [0,1]. This is only true if all the synergies are of the same (negative) sign.

Since the synergy is a measure of the effectiveness of (regional) innovation systems, the assumption of all negative synergies means that those accounted for are the only effective innovation systems. In other words, one can assume that the function $T_{GOT}^{int}/T_{GOT} = f(R_{int}/R)$ is a single value function and can be inverted. Thus, one can measure relative turnover as a function of relative synergy $R_{int}/R = F(T_{GOT}^{int}/T_{GOT})$.

In this study, we have used the TH metaphor as a ladder to estimate the returns of international collaboration on regional economic development. We test the model with Norwegian data for two regions. The investigation into the relation between foreign ownership, international networks, and export for Norwegian firms provides support to the hypothesis that foreign ownership gives the firm a stronger international network, which in turn increases the probability for exports [Menon 2012]. However the situation may differ among regions. In a study of the link between TH synergy and foreign ownership in Hungarian firms, Lengyel and Leydesdorff [2011], for example, found a weak correlation between foreign-owned firms and regional synergy.

**Norwegian geography and economy, some characteristics**

The Norwegian economy features a combination of free-market activities and government interventions. The country is richly endowed with natural resources such as petroleum, hydropower, fish, forest, and minerals. The economy is highly dependent on the petroleum sector, which in 2012 accounted for 23% of the value creation in the country [Ministry of Petroleum and Energy, 2012].

Norway is administratively organized at three levels: the central government (NUTS[6] 1), 19 counties (at the NUTS 3 level) and 430 municipalities at the NUTS 5 level.

A map of the Norwegian counties is given in Figure 2. The capital region surrounding Oslo (county nr. 2) is the most densely populated area of the country. The central government, as well as the major knowledge institutions, are located in Oslo in the southeast. However, the major technical university is located in Trondheim, in the county of Sør-Trøndelag (county nr. 16).

---

[6] Nomenclature des unités territoriales statistiques

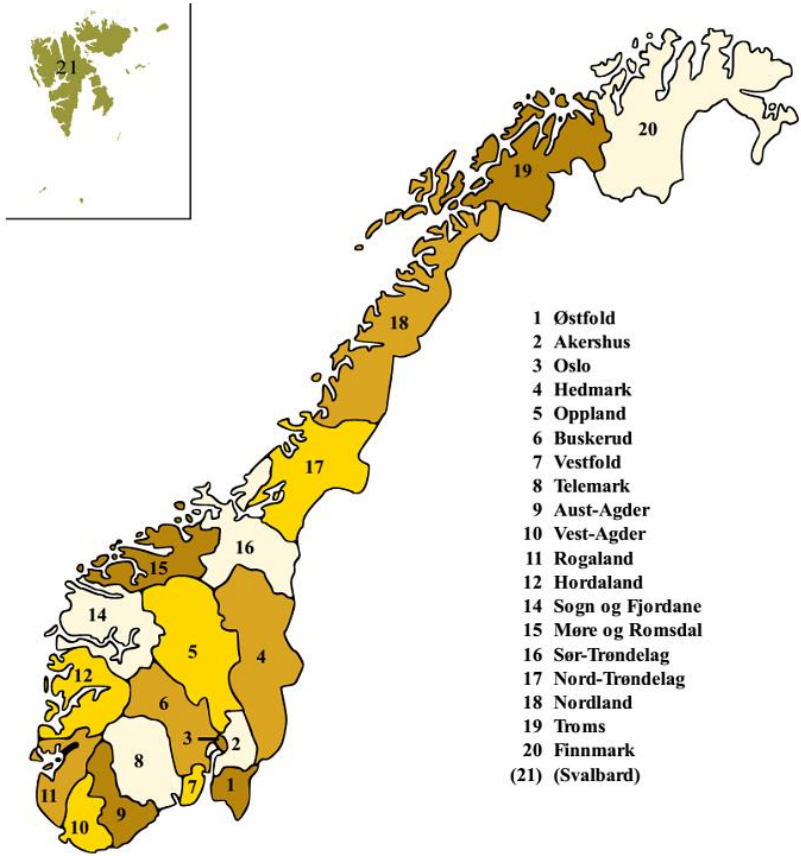

Figure 2 Norwegian counties.

Isaksen [2009] investigated the innovation dynamics of six regional clusters in Norway, which are the main industrial centers. Among others he identified a maritime cluster in Møre og Romsdal, and an instrumentation cluster in Sør-Trøndelag (Trondheim). Further studies of the characteristics of various regional innovation systems in Norway can be found in Asheim and Isaksen [2002], Isaksen and Onsager [2010], Isaksen and Karlsen [2012], and Strand and Leydesdorff [2013]. The analysis of Triple Helix synergy in Norway shows a high level of synergy in Rogaland, Hordaland, Møre og Romsdal, and Nordland. The results for Nordland, however, are not stable over geographical scales.

In this paper we compare two neighboring counties: Møre og Romsdal, with a strong and export-oriented industry and weak knowledge institutions, and Sør-Trøndelag, with strong academic institutions and smaller and more fragmented industry. Møre og Romsdal had up until 2016 three university colleges with less than 700 researchers. Sør-Trøndelag has the main technical university in Norway (NTNU) with close to 6,000 researchers. SINTEF, one of the largest independent research institutes in Scandinavia, with more than 2,000 researchers is also located in Sør-Trøndelag. In 2016, the university college in Ålesund merged with NTNU in Trondheim. This university has its main focus on maritime engineering and business.

The industry in Møre og Romsdal is dominated by the maritime and marine sector. Asheim and Grillitsch [2015] have characterized the county as a peripheral manufacturing region, which performs remarkably well as an economy, despite the lack of strong academic institutions. Frøystad and Nesset [2015] found in their study of maritime suppliers in Møre og Romsdal, that the firms collaborating with global suppliers and customers have higher probability for product innovations, compared to firms collaborating locally. Isaksen [2009] compared the maritime cluster in Møre og Romsdal with the instrumentation cluster in Sør-Trøndelag. The instrumentation cluster is characterized by employees having a high degree of formal education (30% of staff have up to four years of university education and 40% have more than four years). This in contrast to the maritime cluster (20% of staff have up to four years of university education and less than 10% have more than four years). Both clusters are globally competitive and regionally based.

The number of employees in the maritime cluster is an order of magnitude larger than the number of employees in the instrumentation cluster, but these numbers are highly dependent on the inclusion criteria. The numbers of firms as reported by the cluster organizations is 55 firms in Sør-Trøndelag and 200 in Møre og Romsdal [NCE Instrumentation, 2016; GCE Blue Maritime, 2016]. Both clusters are characterized by firms with global value chains. The leading firms in these global value chains are important for external input to the regional cluster firms. Isaksen [2009] reports that both regions have at least two leading firms with a majority of foreign ownership. In Møre og Romsdal the two largest firms (in terms of turnover in 2015) are Vard Group AS, owned by the Italian Fincantiere group, and Rolls-Royce Marine AS, dominated by UK owners. Both are leading firms in the maritime industry.

The largest firms in Sør-Trøndelag are Reitangruppen AS, mainly in groceries, and the publically funded St Olav Hospital. Møre og Romsdal is also the center for fish export from Norway. Fløysand *et al*. [2012] reports that the maritime cluster in Møre og Romsdal is organized bottom-up, whereas most other clusters in Norway are organized top-down. Medium-tech manufacturing firms dominate in this county, whereas in Sør-Trøndelag, small firms in high-tech manufacturing and high-tech services dominate [Strand and Leydesdorff, 2013]. The example of two neighboring counties which are so different makes this a very interesting comparison. A comparison between relevant indicators for the two counties is given in Table 1 below. The TH synergy is compared to R&D expenditure and export income of the region. The table also gives information about the turnover and ownership of the 500 largest firms in each county.

Table 1. Characteristics of the two counties studied in this paper.

| Indicators | Møre og Romsdal | Sør-Trøndelag |
|---|---|---|
| TH Synergy ($T_{GOT}$) | -0.421 bits[7] | -0.204 bits |
| TH *int.* Synergy ($T_{GOT}^{int}$) | -0.24 bits | -0.027 bits |
| R&D expenditure per capita (NIFU-STEP, 2011) | 3.503 NOK[8] | 24.094 NOK |
| Export pr. Employees (Menon, 2012) | 711.000 NOK | 178.000 NOK |
| Population (Menon, 2012) | 249.000 | 287.000 |
| Turnover in the 500 largest firms[9] in 2013 | 170 billion NOK | 185 billion NOK |

---

[7] Bits of Information
[8] Norwegian kroner, 1 NOK= 0.118 Euro or 0.134 $
[9] Turnover based on data from Purehelp.no

| Foreign owned firms[10] amongst the 500 largest firms in 2013 | 44 | 39 |
| Turnover in foreign firms compared to domestic | 24% | 9% |
| $(R_{int}/R)/(T_{GOT}^{int}/T_{GOT})$ | 0.25 | 0.68 |

**Results**

We performed the calculations for the Sør-Trøndelag and Møre og Romsdal counties by focusing on the 500 firms with highest turnover in both counties, respectively. There are 39 foreign-owned firms in Sør-Trøndelag and 44 foreign-owned firms in Møre og Romsdal among these 500. As indicated in Table 1 the summary turnover is approximately 185 billion NOK and 170 billion NOK, respectively. The distribution of domestic and foreign firms in the two counties is similar. Chi-square analysis of the distributions between domestic and foreign owned firms in the two counties is 0,328 ($p$ = 0.647). This implies that these distributions cannot be considered independent. This fact can be attributed to strong collaboration with global suppliers and customers which impose some unification.

*a. Sør-Trøndelag*

The total Sør-Trøndelag county ternary synergy $T_{GOT}$ estimated for all 500 county firms is –0.204 (in bits of information). The part of the synergy generated by foreign-owned firms ($T_{GOT}^{int}$ = –0.027 bits of information). The ratio of internationally generated turnover $R_{int}$ to summary turnover $R$ *vs.* internationally generated synergy surplus $T_{GOT}^{int}$ to total synergy $T_{GOT}$ is 0.68. This ratio can be considered as a measure of relative international synergy surplus effectiveness in terms of relative turnover surplus.

---

[10] Ownership data from Proff.no

The distribution of foreign-owned firms by technology groups follows the distribution of the domestically owned firms (Figure 3). The main spheres of activity of domestically owned firms correspond to the second (manufacturing, mining and quarrying, and other industry), third (construction), fourth (wholesale and retail trade, transportation and storage, accommodation, and food service activities), and eighth (professional, scientific, technical, administration, and support service activities) technology groups with a focus on trade, transportation, and food service activities, whereas foreign-owned firms are mostly engaged in the activities corresponding to second, fourth, and eighth technology groups with an emphasis on manufacturing, mining, and other industrial applications.

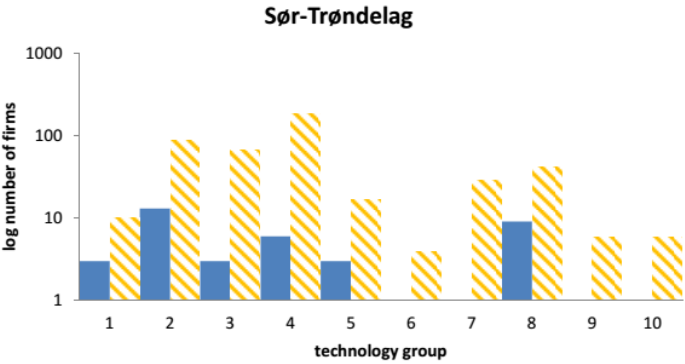

Figure 3. Log of the total number of firms and number of foreign-owned firms with respect to technology groups for Sør-Trøndelag county

Due to largely coinciding spheres of domestically and foreign-owned firms' activities, international collaboration does not bring additional substantial diversification to the regional economy and regional exports, and we find a statistically and economically important relationship between export growth and income growth [Lewer and Berg 2003; Hidalgo and Hausmann 2009]. At the same time, boosting synergy in the international dimension demands additional efforts and expenses so that at initial stages with a comparatively low level of internationally owned firms the return is not so substantial. The specific role of foreign-owned companies is that they can be considered as a form of foreign investment, which has an effect on knowledge transfer, information sharing, technology spillover, and the development of human capital.

Furthermore, one should consider the effect of foreign-owned companies on the development of regional innovation systems, since international collaboration brings an additional dimension to university-industry-government relations. From the literature related to the cluster theories, the concept of "local buzz-global pipelines" is well known [Bathelt et al., 2004]. The local knowledge flows are characterized by informal exchanges of applied knowledge related to ongoing projects; this "local buzz" is highly efficient given that the actors are co-located. The global knowledge flow relates to contact with customers in global markets, this "global pipeline" brings state of the art knowledge from global markets back to the local cluster. This means that local or national knowledge institutions may be bypassed by these global pipelines if they do not interact with local industries.

*b. Møre og Romsdal*

In Møre og Romsdal, the total ternary synergy $T_{GOT}$ estimated for all 500 county firms equals –0.421 bits of information. This value twice exceeds the value of Sør-Trøndelag. A large part of total synergy is generated by foreign-owned firms: $T_{GOT}^{int} = $ –0.396 bits. The ratio of internationally generated turnover $R_{int}$ to total turnover $R$ *vs.* the ratio of internationally generated synergy surplus $T_{GOT}^{int}$ to total synergy $T_{GOT}$ is 0.42. This value is lower than corresponding value for Sør-Trøndelag county. In other words the effect of relative international synergy increment in terms of relative international turnover increment in Møre og Romsdal is less accentuated than in Sør-Trøndelag. Figure 4 presents the logarithm of the distribution of domestically owned and foreign-owned firms by technology groups.

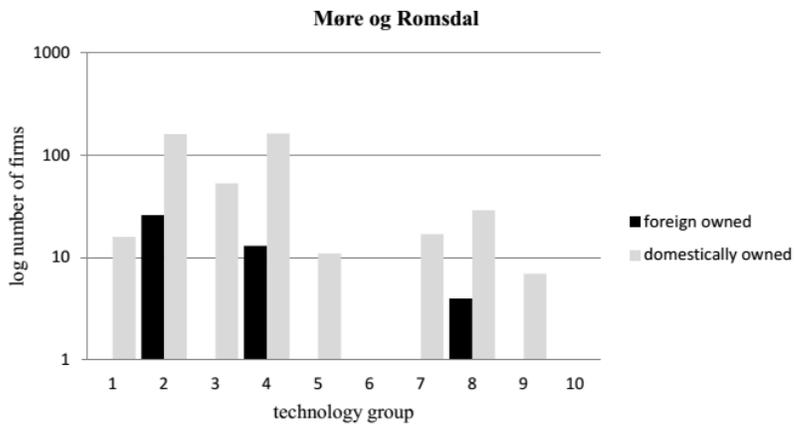

Figure 4. Log of the total number of firms and number of foreign-owned firms with respect to technology groups for Møre og Romsdal county

**Discussion**

When comparing results for the two counties the difference is striking. Most of the synergy in Møre og Romsdal is attributed to the firms with foreign ownership (-0.24/-0.421 = 57.1%), whereas foreign-owned firms account for only a small part of the synergy in Sør-Trøndelag (-0.027/-0.204 = 13.2%). Based on previous calculations using data for all Norwegian firms, one would expect the production- and export-oriented county of Møre og Romsdal to demonstrate other characteristics than the more science- and knowledge-oriented county of Sør-Tøndelag [Strand and Leydesdorff, 2013]. However, the different results (Figure 7) are rather extreme.

Based on our knowledge of the two counties we would expect foreign ownership in low- and medium-tech industries, such as shipbuilding and ship equipment production. Generally, global integrated value chains characterize the maritime offshore industry. The firms are mature, well-established, and large. In Sør-Trøndelag, we would expect a higher number of new, small high-tech companies in their growing phase. These small firms will have relatively small turnover compared to the more mature firms in Møre og Romsdal. We would expect that an analysis of a larger share of firms for both counties would dampen the effect for Møre og Romsdal and increase the effect of foreign ownership on synergy in Sør-Trøndelag. (However, inclusion of ownership data requires two separate databases to be matched manually for each firm.)

This said, how can it be that a county like Sør-Trøndelag with ten times the number of academics and researchers compared with Møre og Romsdal, does not demonstrate the same level of Triple Helix synergy? Can this be a sign of fragmentation, as reported by Onsager et al. [2010] or "parallel worlds", as reported by OECD [2006], in the situation for the academic institutions in Sør-Trøndelag? Is it so that researchers at these institutions prefer career-relevant academic research in favor of working together with the industry on more applied (and perhaps less easily publishable) problems? Alternatively, do the knowledge resources in Sør-Trøndelag act as knowledge bank for the national industry and can the results of knowledge transfer from these institutions only be detected in industrial regions where these results are used?

The industrial structure as such may also affect these results. For example, a large offshore construction vessel with a price tag of several billons NOK (Norwegian kroner) needs a huge number of regional suppliers compared with a high-tech firm producing a small series of advanced instrumentations. But an alternative would be provided by the previously mentioned local buzz-global pipeline hypotheses, where industry-relevant knowledge flows directly from global customers to the local cluster firms. It has previously been suggested [Strand and Leydesdorff, 2013] that the national knowledge institutions are bypassed if not relevant for the knowledge flow. For small firms located in a regional cluster where the central role is played by internationally owned firms, as in Møre og Romsdal, this may provide opportunities for "piggy-backing" where the small firms follow the internationally leading firms. Using this mechanism, small firms with few resources may still be able to obtain international contracts and generate export income on global markets.

The role of the knowledge institutions (academia) is very interesting because of the merger between a strong academic partner (NTNU) and a more applied and industry-focused regional university college in Ålesund. According to GVC and cluster literature, the knowledge institutions play a central role in the various improvement processes caused by competitive pressure and new knowledge. Upgrading takes place both vertically along the value chains and horizontally among the cluster firms. Interaction between the strong academic institutions in Sør-Trøndelag and the strong industry in Møre og Romsdal may enhance the cluster and value-chain improvements. Likewise, there is a danger that the industry-focused knowledge institution embedded in the maritime cluster will increasingly be directed towards more academic and less applied research.

National and regional governments have also an important role in developing conditions that are attractive to the leading firms. Internationally owned firms are more likely to locate their R&D facilities in regions and countries with favorable research funding and strong knowledge institutions. Governmental policies should encourage the interaction between firms and knowledge institutions in order to enhance the understanding of challenges faced by industry that can perhaps be solved with new knowledge from academia.

**Conclusion**

By using ownership data for the 500 largest firms (in terms of turnover) in two neighboring counties, we showed that foreign ownership has a strong effect on synergy. From previous studies we know that the level of triple-helix synergy is higher in Møre og Romsdal, compared to Sør-Trøndelag. However, the ratio $(R^{int}/R)/(T^{int}_{GOT}/T_{GOT})$ in Sør-Trøndelag (0.68) is approximately half times greater than that in Møre og Romsdal (0.42). In other words, the relative international synergy increment in the efficiency, in terms of relative turnover, is higher in Sør-Trøndelag than in Møre og Romsdal. This result can be explained by the larger R&D expenditure per capita in Sør-Trøndelag. Our results suggest that it is easier to improve the TH synergy in Sør-Trøndelag than in Møre og Romsdal, since the available potential has not yet been used.

From a methodological perspective, our results show that one can link the abstract concept of Triple Helix synergy, measured in bits of information, to more familiar economic terms like turnover.  Variations in industry structure as well as maturity of the industry between the two counties may explain the strong effects that were detected. We expect that the inclusion of more firms will dampen these effects; however, this remains to be shown.

What can one do to enhance synergy in a region? Answers from Triple Helix theory, cluster theory, and research on GVC all point to the role of increased interactions. From the perspective of Triple Helix theory, the interactions are between the actors in industry, academia and government. Cluster theory points to interactions and localized learning among the firms in a cluster, whereas GVC emphasizes the role of interaction between the firms and the global customer. The findings from the three streams of research and the results from the above calculations point to the central role of the internationally owned firms in the clusters. Internationally owned firms seem to be a key element for enhancing synergy in a region. While this role was disruptive in Hungary as a country with mainly a local industry, it can be appreciated in Norway as a country with an open economy. However, further research is needed for clarifying the various aspects of the roles of lead firms. The observed effect may be due to the relations between the knowledge institutions, the government, and other firms in each specific region.

**Acknowledgements**

This paper was presented at the conference Cross-Border Markets of Goods and Services: Issues for Research, held by the Far Eastern Federal University in Vladivostok, Russia, May 27-28, 2015.  We acknowledge comments and feedback from several colleagues in the discussion. We thank two anonymous reviewers for insightful comments and suggestions for improvements of the manuscript. Inga Ivanova acknowledges support within the framework of the Basic Research Program at the National Research University Higher School of Economics (NRU HSE) and within the framework of a subsidy by the Russian Academic Excellence Project '5-100'.

**Conflict of interests**

The authors declare that they have no conflicts of interest.

**References**

Abramson, N. (1963). *Information Theory and Coding*. New York, etc.: McGraw Hill.

Ashby, W. R. (1964). Constraint analysis of many-dimensional relations. *General Systems Yearbook, 9*, 99-105.

Asheim, B. T., & Grillitsch, M. (2015). *Smart specialisation: Sources for new path development in peripheral manufacturing region.* In Strand, Ø., Nesset, E., Yndestad, H. (Eds.), Fragmentering eller mobilisering? Regional utvikling i Nordvest, 87-113. Ålesund, Norway: Forlag1

Asheim, B.T., & Isaksen, A. (2002), Regional Innovation Systems: The Integration of Local 'Sticky' and Global 'Ubiquitous' Knowledge, Journal of Technology Transfer, Vol. 27, No. 1, pp. 77-86.

Bathelt, H., Malmberg, A., Maskell, P. (2004). Clusters and knowledge: local buzz, global pipelines and the process of knowledge creation. *Progress in Human Geography*, 28(1), 31-56.

Blau, P. & Schoenherr, R. (1971). *The Structure of Organizations*. New York: Basic Books, 435 pp.

GCE Blue Maritime. (2016). http://www.bluemaritimecluster.no/ Accessed 15.04.2016

Cooke, P. (1992) Regional innovation systems: competitive regulation in the new Europe, *Geoforum 23*, pp. 365-382.

Cooke, P., Memedovic, O. (2003). Strategies for regional innovation systems: Learning transfer and applications. http://www.unido.org/fileadmin/import/11898_June2003_CookePaperRegional_Innovation_Systems.3.pdf%20Accessed 2008.04.2016

Eurostat (2008). *NACE Rev.2 – Statistical classification of economic activities in the European Community*, Luxemburg: Office for Official Publications of the European Communities, http://gac.gov.ge/files/safasuri/2_3.pdf

Fløysand, A., Jakobsen, S.E., Bjarnar, O. (2012). The dynamism of clustering: Interweaving material and discursive processes. *Geoforum,* 43, 948-958.

Freeman, C. (1987). Technology policy and economic performance: Lessons from Japan. London: Pinter.

Frøystad, M. K., & Nesset, E. (2015). Geographical sources of innovation for upstream companies in a regional maritime cluster. In Strand, Ø., Nesset, E., Yndestad, H. (Eds.), Fragmentering eller mobilisering? Regional utvikling i Nordvest, 87-113. Ålesund, Norway: Forlag1

Gereffi, G., (1994). The Organisation of Buyer-Driven Global Commodity Chains: How US Retailers Shape Overseas Production Networks. In G. Gereffi, and M. Korzeniewicz (Eds), Commodity Chains and Global Capitalism. Westport, CT: Praeger.

Gereffi, G., Lee, J. (2012). Why the world suddenly cares about global supply chains. *Journal of Supply Chain Management*, 48(3), 24-32

Gereffi, G., Lee, J. (2016). Economic and Social upgrading in Global Value Chains and Industrial Clusters: Why Governance Matters, *Journal of Business Ethics*, 133, 25-38.

Hidalgo, C. A. and Hausmann, R. (2009). The building blocks of economic complexity. *Proceedings of the National Academy of Sciences 106*(26). 10570–75.

Humphrey, J., Schmitz, (2002). How does insertion in global value chains affect upgrading in industrial clusters? *Regional Studies*, *9*(36), 1017-1027.

Isaksen, A. (2009). Innovation dynamics of global competitive regional clusters: the case of the Norwegian center of expertise, *Reg. Stud.* 43 (9), 1155–1166.

Isaksen, A. & Onsager, K. (2010). Regions, network and innovation performance: the case of knowledge-intensive industries in Norway, *Eur. Urban Reg. Stud.* 17 (3), 227–243.

Isaksen, A., & Karlsen, J. (2012). Can small regions construct regional advanteges? The case of four Norwegian regions. *European Urban and Regional Studies* 20(2), 245-257.

Krippendorff, K. (2009). Information of Interactions in Complex Systems. *International Journal of General Systems, 38*(6), 669-680.

Lengyel, B., Leydesdorff, L. (2011). Regional Innovation Systems in Hungary: The failing Synergy at the National Level. *Regional Studies 45*(5) (2011), 677-693.

Leydesdorff, L., Dolfsma, W., & Van der Panne, G. (2006). Measuring the Knowledge Base of an Economy in terms of Triple -Helix Relations among 'Technology, Organization, and Territory' *Research Policy, 35* (2), 181-199.

Leydesdorff, L., & Etzkowitz, H. (1996). Emergence of a Triple Helix of University – Industry Government relations, *Science and Public Policy* 23, 279-286.

Leydesdorff, L. & Fritsch, M. (2006). Measuring the knowledge base of regional innovation systems in Germany in terms of a Triple Helix dynamics, *Research Policy 35* 1538-1553.

Leydesdorff, L. & Ivanova, I, (2014). Mutual Redundancies in Inter-human Communication Systems. Steps Towards a Calculus of Processing Meaning. *Journal of the Association for Information Science and Technology 65*(2), 386-399.

Leydesdorff, L., Perevodchikov, E., & Uvarov, A, (2015). Measuring Triple-Helix Synergy in the Russian Innovation Systems at Regional, Provincial, and National Levels, *Journal of the Association of Information Science and Technology* 66(5), 1001-1016.

Leydesdorff, L. & Strand, Ø (2013). The Swedish System of Innovation: Regional Synergies in a Knowledge-Based Economy. *Journal of the American Society for Information Science and Technology* 64(9) 1890-1902; DOI: 10.1002/asi.22895.

Lewer, J. J., & Berg, H. V. den (2003). How Large Is International Trade's Effect on Economic Growth? *Journal of Economic Surveys*, *17*(3). 363–96.

Lundquist, K. J., & Trippl, M. (2013). Distance, Proximity and Types of Cross-border Innovation Systems: A Conceptual Analysis. *Regional Studies, 47*(3), 450-460.

Lundvall, B.-Å. (1988). Innovation as an interactive process: from user–producer interaction to the national innovation systems. In: Dosi, G., Freeman, C., Nelson, R.R., Silverberg, G., Soete, L. (Eds.), *Technical Change and Economic Theory*. Pinter, London.

Lundvall, B.-E. (1992). Introduction. In: Lundvall, B.-E. (Ed.) *National systems of innovation: Towards a theory of innovation and interactive learning*. London: Pinter.

Maskell, P. and Malmberg, A. (1999a). The competitiveness of firms and regions: 'ubiquitification' and the importance of localized learning. *European Urban and Regional Studies 6*, 9-25.

Maskell, P. and Malmberg, A. (1999b). Localised learning and industrial competitiveness. Cambridge Journal of Economics 23, 167-185.

Ministry of Petroleum and Energy (2012). Norways's oil history in 5 minutes. Retrieved from https://www.regjeringen.no/en/topics/energy/oil-and-gas/norways-oil-history-in-5-minutes/id440538/. Accessed 10.05.2016

Menon (2012). *Eksport fra norske regioner-Hvorfor så store forskjeller* (Norwegian) (Export from Norwegian regions- why are there so large differences?), Menon-publication nr.

2/2012. Retrieved from http://menon.no/upload/2012/02/27/eksport-fra-norske-regioner-2009.pdf    Accessed 10.05.2016

NCE Instrumentation. (2016). http://ncei.no/english/ Accessed 15.04.2016

OECD (2006). Supporting the contribution of higher education institutions to regional development, peer review report: Trøndelag (Mid-Norwegian Region), Norway 2006. retrieved April 15, 2011 from: http://www.oecd.org/dataoecd/8/2/37089380.pdf. 4

Onsager, K., Aslesen, H.W., Gundersen, F., Isaksen, A., Langeland, O. (2010). *City Regions, advantages and innovation*, NIBR report 2010(5), Oslo, 2010. Retrieved from: http://www.nibr.no/uploads/publications/f1fcfca6cdf727353a4f99114e1b6f9e.pdf. Accessed 10.05.2016

Strand, Ø., & Leydesdorff, L. (2013). Where may synergy be Indicated in the Norwegian Innovation System? Triple Helix Relations among Technology, Organization, and Geography, *Technological Forecasting and Social Change 80*(3), 471-484.

Theil, H. (1972). *Statistical Decomposition Analysis*. Amsterdam/ London: North-Holland.

Van den Broek, J. & Smulder, H. (2013). The evolution of a cross - border regional innovation system: An institutional perspective. Regional Studies Association European Conference, Tampere.  http://www.regionalstudies.org/uploads/Van_Den_Broek_Smulders.pdf Accessed 10.05.2016

Yeung, R.W. (2008). *Information theory and network coding.* New York, NY: *Springer.*

Appendix

Table 2. Correspondence between high level aggregation of ISIC/NACE categories and two digit NACE Rev. 2 codes

| ISIC Rev. 4/NACE Rev. 2 high-level aggregation | ISIC Rev. 4/NACE Rev. 2 sections | NACE Rev.2 two digit codes | Description |
|---|---|---|---|
| 1 | A | 01-03 | Agriculture, forestry and fishing |
| 2 | B, C, D, E | 05-39 | Manufacturing, mining and quarrying, and other industry |
| 3 | F | 41-43 | Construction |
| 4 | G, H, I | 45-56 | Wholesale and retail trade, transportation and storage, accommodation and food service activities |
| 5 | J | 58-63 | Information and communication |
| 6 | K | 64-66 | Financial and insurance activities |
| 7 | L | 68 | Real estate activities |
| 8 | M, N | 69-82 | Professional, scientific, technical, administration and support service activities |
| 9 | O, P, Q | 84-88 | Public administration, defense, education, human health and social work activities |
| 10 | R, S, T, U | 90-99 | Other services |